\documentclass{kapproc} 
\usepackage[dvips]{graphicx}
\begin{document}

\articletitle{The Collinder 69 cluster in the context of the  Lambda Orionis SFR}
\articlesubtitle{An Initial Mass Function Down to the Substellar Domain}

\author{David Barrado y Navascu\'es}
\affil{LAEFF-INTA, Madrid (SPAIN)}
\email{barrado@laeff.esa.es}

\author{John R. Stauffer}
\affil{IPAC, California Institute of Technology (USA)}
\email{stauffer@ipac.caltech.edu}

\author{Jerome Bouvier}
\affil{Laboratoire d'Astrophysique, Observatoire de Grenoble (FRANCE)}
\email{Jerome.Bouvier@obs.ujf-grenoble.fr}

\author{Ray Jayawardhana}
\affil{University of Toronto  (CANADA)}
\email{rayjay@astro.utoronto.ca}

\begin{abstract}
The Lambda Orionis Star Forming Region is a complex structure
which includes the Col~69 (Lambda Orionis) cluster
 and the B30 \& B35 dark clouds.
We have collected deep optical photometry and spectroscopy
in the central cluster of the SFR (Col~69), and combined
with 2MASS IR data, in order 
to derive the Initial Mass Function of the cluster,
in the range 50-0.02 $M_\odot$. In addition, we have studied
the H$\alpha$ and lithium equivalent widths, 
 and the optical-infrared photometry,
to derive an age  (5$\pm$2 Myr) for Col~69, and to compare
these properties to those of B30 \& B35 members.
\end{abstract}


\begin{figure}   
  \includegraphics[width=\columnwidth]{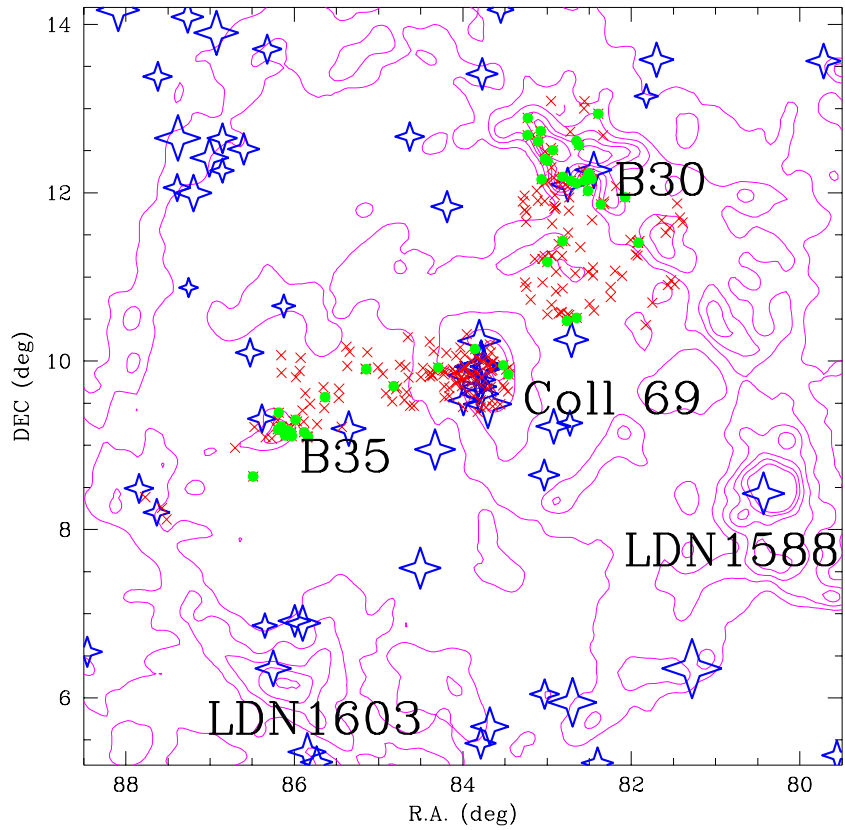}
  \caption{
IRAS map at 100 $\mu$ of the Lambda Orionis Star Forming Region
(contour levels in purple). B stars are displayed as blue four points
 stars (size related to brightness). Red crosses and green, solid circles
represent  stars  listed in D\&M. In the case of the 
green circles, they have an excess in the H$\alpha$ emission (see
Figure 11). We have labeled the location of several stellar associations
and dark clouds. Note the concentration of B stars in the Nort-East edge
of the ring, with might indicate the presence of another cluster.
}
\end{figure}

\section{Introduction}

One of the most  prominent Star Forming Regions (SFRs)
 in Orion, albeit not very well studied,
 is the Head of Orion, the 
Lambda Orionis Star Forming Region (hereafter LOSFR).
This area, about 50 sq.deg., is dominated by the O8 III star
$\lambda$$^1$ Orionis.
This star is located at   the center of a  ring of dust and molecules,
 which  encloses  the S264 H{\sc II} region, a large number of IRAS sources, 
the LDB1588 \& LDN1603 and  Barnard 30 and Barnard 35 (B30 \& B35)
 dark clouds, and the Lambda Orionis (also called Collinder 69) cluster
--which is located just  around the central star.
Figure 1 shows this region, and includes the emission at 100 $\mu$
(from IRAS) and the distribution of stars of B spectral type.

Several groups have   focused on different aspects of the LOSFR,
such as   the photometric properties of the 
high mass members (Murdin \& Penston 1977),
a H$\alpha$ survey (Duerr, Imhoff \& Lada 1982),
the analysis of the IRAS data (Zhang et al. 1989)
 and a  photometric and spectroscopic search 
(Dolan \& Mathieu 1999,  2001, 2002, D\&M hereafter).
All this wealth of data has shown that young stars  
are clustered around
 the central part of the LOSFR (the Col~69 cluster)
 and the dark clouds B30 \& B35, whereas a large number of IRAS sources
 are associated to the much denser clouds LDN1588 \& LDN1603.
The distance to this  region is in between 380$\pm$30 pc, as derived 
by Hipparcos, and 450$\pm$50 pc (D\&M).
D\&M also derived an age of about 6 Myr, which 
correspond to the turn-off age for the massive stars
(4 Myr in the case of Murdin \& Penston 1977).
The region  is also characterized by the low reddening in the central area,
 E$(B-V)$=0.12 (Diplas \& Savage 1994). 

This SFR, due to its properties, is an excellent field to study 
different aspects of the stellar and substellar evolution, from
brown dwarfs hunting to studies of the Initial Mass Function (IMF)
and the effect  of the environment, as well as different stellar
properties. In this paper we focus on  the Col~69 cluster and compare 
some of its properties to those of  the B30 \& B35 clouds.

\begin{figure}   
  \includegraphics[width=\columnwidth]{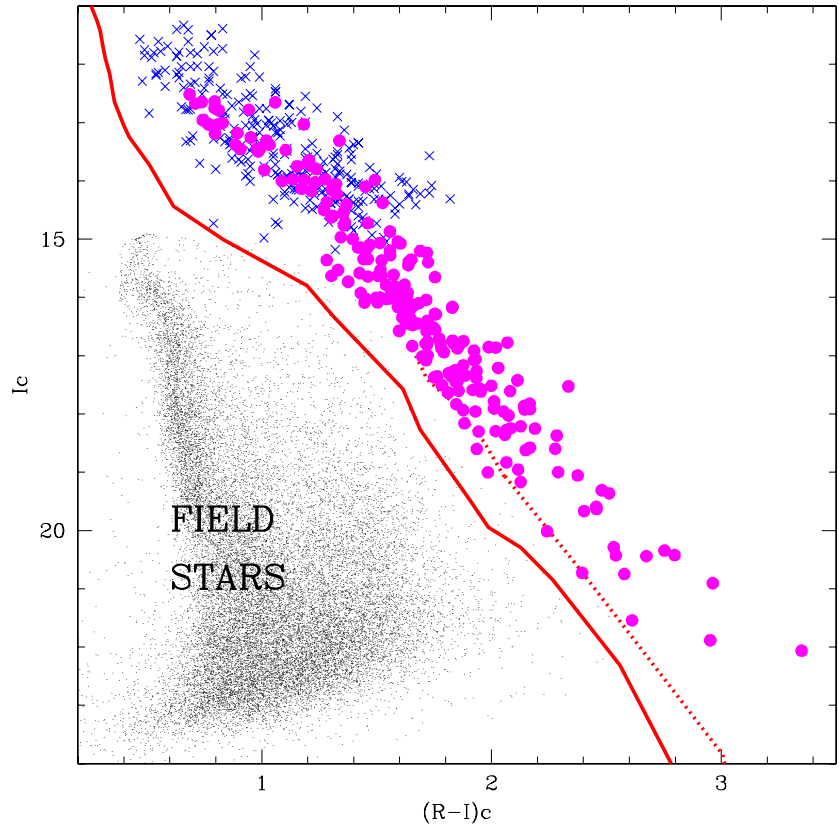}
  \caption{
Our initial deep search in the $R,I$ filters, based on CFHT/12K data
and centered around the $\lambda$$^1$ Orionis star.
The  selection of candidate members appears as magenta circles, whereas
data from D\&M are displayed as blue circles. A 5 Myr Isochrone
and an empirical MS are included as red dotted and solid lines, respectively.
}
\end{figure}

\section{Searching for additional low mass  members of Col~69}

\subsection{Deep optical photometry}

We have conducted a deep optical survey with the CFHT 
12K mosaic camera (hereafter CFHT/12K), which covers 
a field of view  of 42$\times$28 arcmin,
 and the $R_C$ and $I_C$ filters (Cousins). We centered 
the search on the star $\lambda$$^1$ Orionis. The details
can be found in Barrado y Navascu\'es et al. (2004a).
The limiting and completeness magnitudes are $I_C$(lim)=24.0
and $I_C$(compl)=22.75 mag. For cluster members, the completeness 
is achieved at 20.2 mag, due to the limitations imposed by the 
$R_C$ filter.

Figure 2 displays an optical Color-Magnitude (CMD) with our photometric 
data. In the figure, we have included  the field stars as
black dots, our selection of candidate members as purple solid circles
and previously known members of the SFR as blue crosses 
(from D\&M). We have incorporated 
a Zero Age Main Sequence --red solid line-- and a 5 Myr
isochrone from Baraffe et al. (1998) --red dotted line.
They were shifted to the assumed distance of the Col~69
cluster, 400 pc --(m-M)$_0$=8.010-- and its color excess
E($B-V$)=0.12, which translates into E$(R-I)$=0.084
and $A_I$=0.223 (following Rieke \& Lebofsky 1985).
We have selected 170 candidate members with magnitudes in the interval
$I_C$=12.52--22.06 mag.
A more detailed CMD can be found in Figure 3, 
where we represent our selected candidate members
(purple symbols) with those selected by D\&M
(blue crosses for the area around the star $\lambda$$^1$ Orionis,
green crosses for those closer to B30 and B35). In all cases, large
circles denote objects with H$\alpha$ emission well above the saturation
limit (see section 3.1 and Figure 11).
In principle, based on Baraffe   et al. (2003) models (the so called COND
models), we would have reached 0.01 $M_\odot$ if the faintest object is 
indeed a bona fide member  of the cluster.

\begin{figure}   
  \includegraphics[width=\columnwidth]{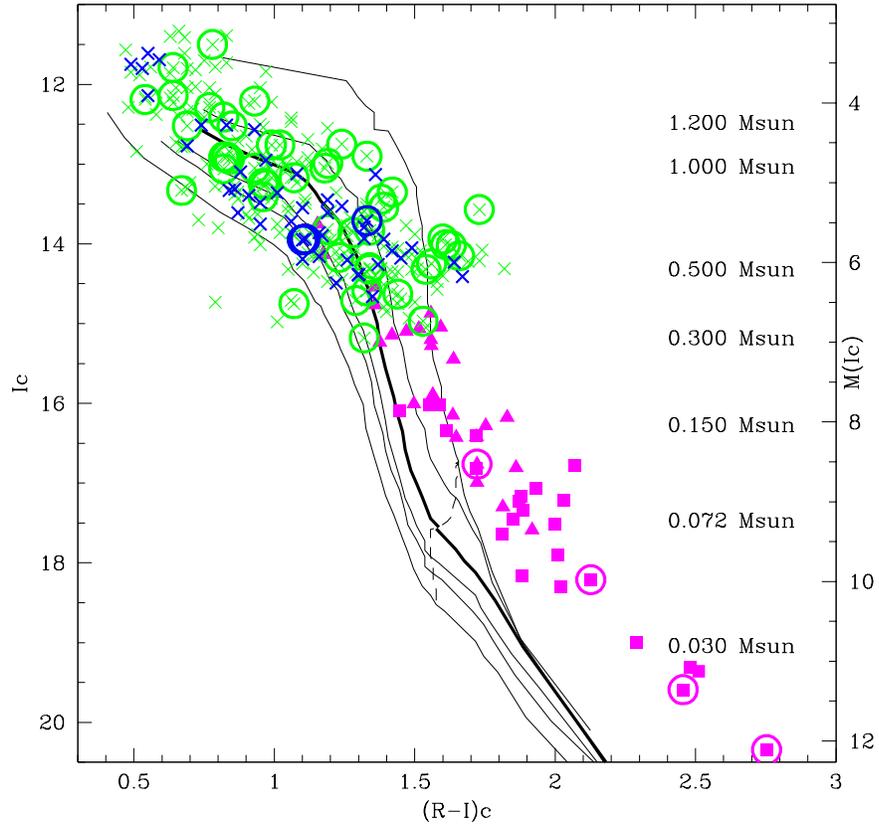}
  \caption{ 
The selection of possible members located in the LOSFR. 
Blue (from D\&M) and purple (our deeper data)
symbols correspond to the Collinder 69 open cluster.
Members associated to the B35 and B30 clouds are shown as green crosses
(also from D\&M).
Big, overlapping  open circles correspond to possible Classical TTauri stars
and substellar analogs (objects having an H$\alpha$ excess, see Figure 11).
Masses, derived using the models by Baraffe et al. (1998), are labeled 
in the right-hand side of the diagram.
}
\end{figure}

\begin{figure}   
  \includegraphics[width=\columnwidth]{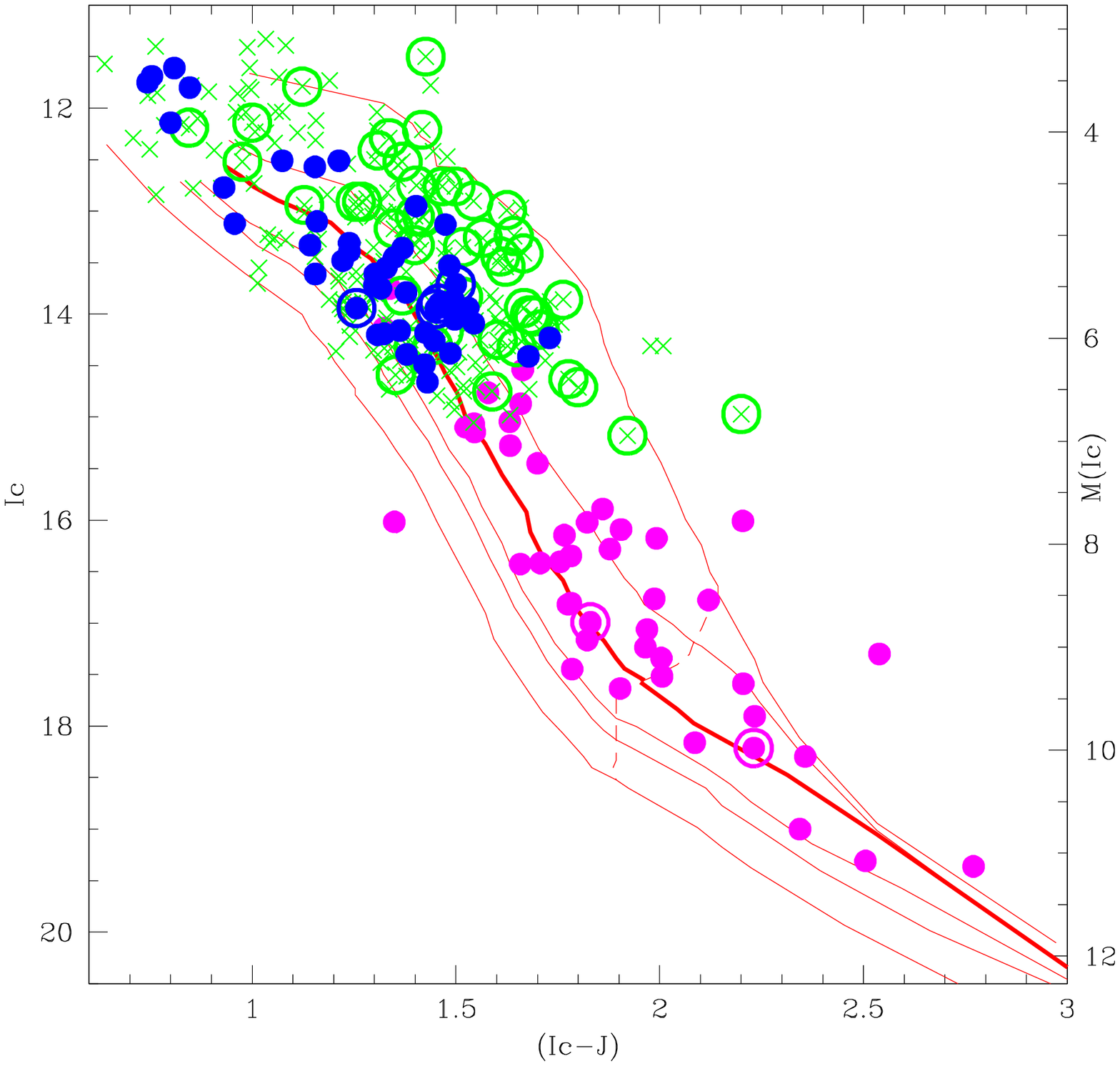}
  \caption{
Optical-Infrared Color-Magnitude Diagram of Coll~69 
(blue and purple solid circles)
and B30 \& B35 (green crosses). The isochrones are those of Baraffe et al. 
(1998), for ages of 1, 3, 5 (highlighted as a bold line), 8, 10 \& 16 Myr.
The dashed, red line is the borderline between stars and brown dwarfs.
Symbols as in Figure 3. 
}
\end{figure}

\begin{figure}   
  \includegraphics[width=\columnwidth]{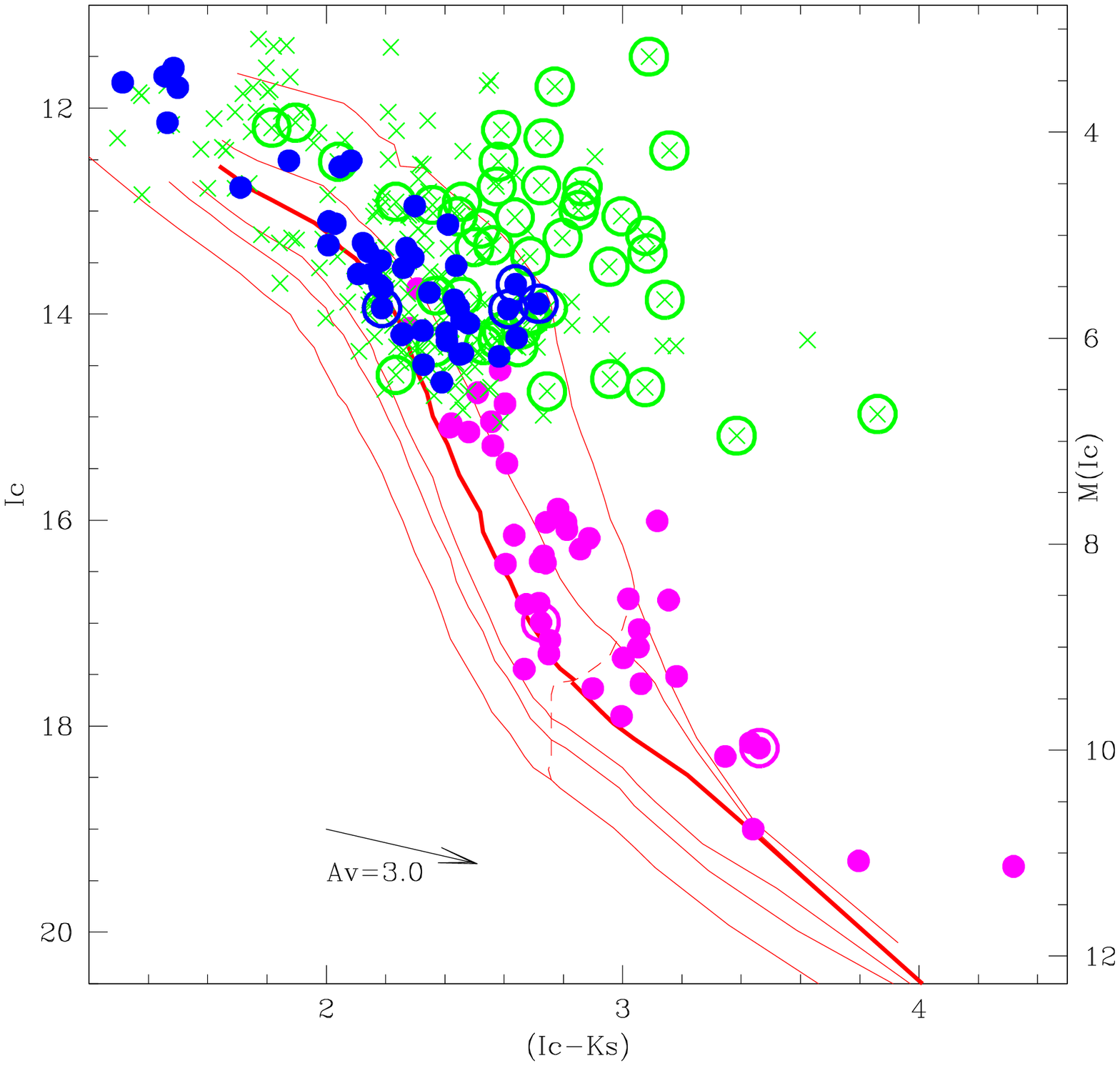}
  \caption{
Optical-Infrared Color-Magnitude Diagram of Coll~69
 (blue and purple solid circles)
and B30 \& B35 (green crosses). The isochrones are those of Baraffe et al. 
(1998), for ages of 1, 3, 5 (highlighted as a bold line), 8, 10 \& 16 Myr.
The dashed, red line is the borderline between stars and brown dwarfs.
Symbols as in Figure 3. 
}
\end{figure}

\begin{figure}   
  \includegraphics[width=\columnwidth]{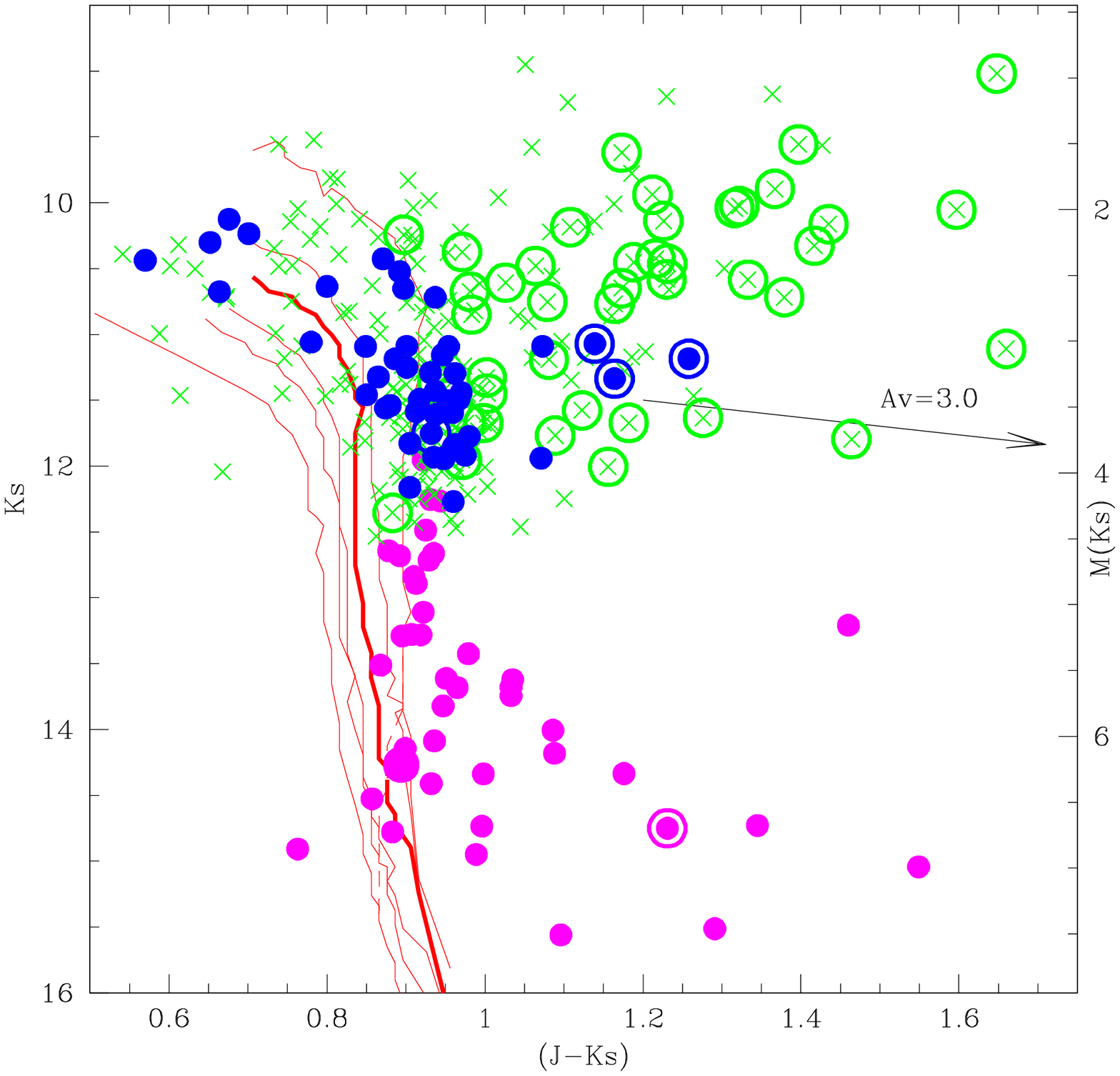}
  \caption{
Infrared Color-Magnitude Diagram of Coll~69 (blue and purple solid circles)
and B30 \& B35 (green crosses). The isochrones are those of Baraffe et al. 
(1998), for ages of 1, 3, 5 (highlighted as a bold line), 8, 10 \& 16 Myr.
The dashed, red line is the borderline between stars and brown dwarfs.
Symbols as in Figure 3. 
}
\end{figure}

\begin{figure}   
  \includegraphics[width=\columnwidth]{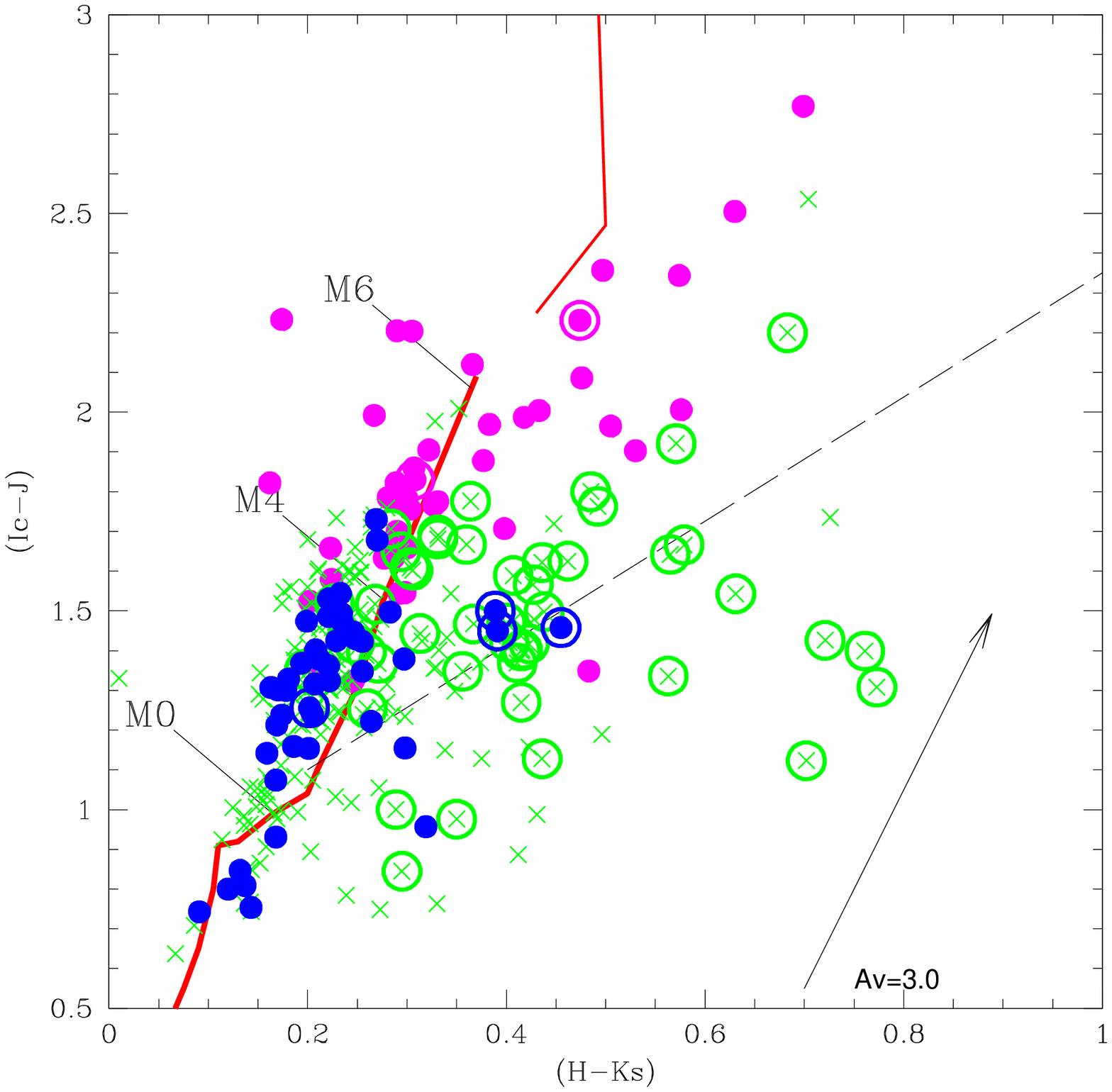}
  \caption{
Optical-Infrared Color-Color Diagram of Coll~69 (blue and purple solid circles)
and B30 \& B35 (green crosses). 
The red line is the locus for Main Sequence Stars  
(from Bessell \& Brett 1988; Kirkpatrick et al$.$
 2000; Leggett et al. 2001). The black, dashed line is the locus
for  Classical TTauri stars
 (Meyer et al$.$ 1997 and Barrado y Navascu\'es et al. 2003).  
Symbols as in Figure 3. 
}
\end{figure}

\begin{figure}   
  \includegraphics[width=\columnwidth]{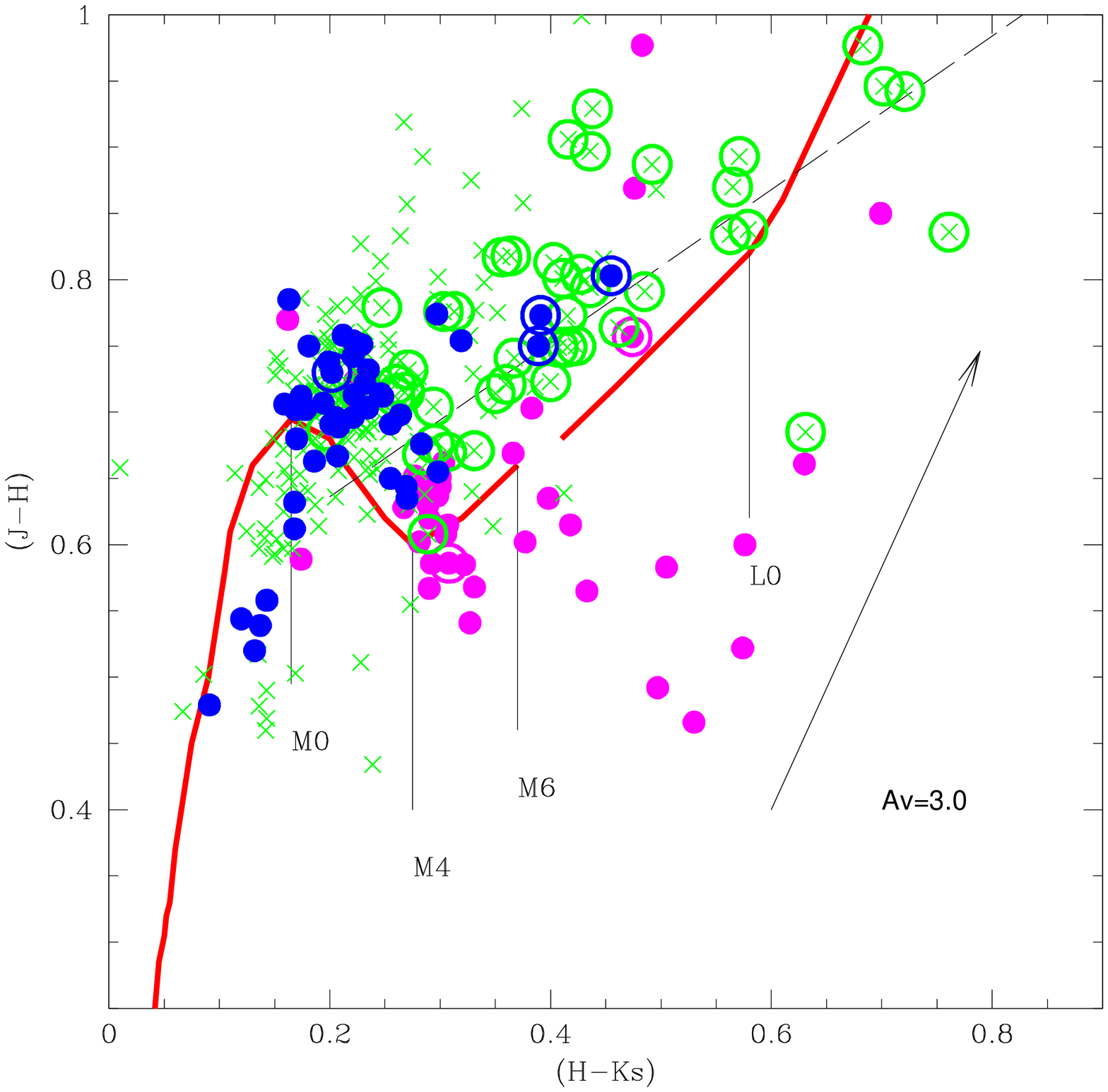}
  \caption{
Infrared Color-Color Diagram of Coll~69 (blue and purple solid circles)
and B30 \& B35 (green crosses). 
The red line is the locus for Main Sequence Stars  
(from Bessell \& Brett 1988; Kirkpatrick et al$.$
 2000; Leggett et al. 2001). The black, dashed line is the locus
for  Classical TTauri stars
 (Meyer et al$.$ 1997 and Barrado y Navascu\'es et al. 2003).  
Symbols as in Figure 3. 
}
\end{figure}

\subsection{Follow-up: 2MASS photometry}

We have mined the 2MASS database (All Sky release, Cutri et al. 2003)
in order to complement our optical data with near infrared photometry.
Due to the limiting magnitude of that survey, we were only able to
get these data for objects brighter than about $I_C$=19.1 mag.
In any case, we have used these data to classify the objects 
initial selected as candidate members into
 probable members and possible members, and
probable and possible non-members. See Table 2 of 
Barrado y Navascu\'es et al. (2004a).

We have also searched for the 2MASS counterparts 
of the D\&M candidate members.
Figure 4-6 includes several  CMD using the optical and infrared 
data. Symbols are as in Figure 3.
The diagrams suggest that those D\&M candidates clustered
around the B30 and B35 darks clouds (green crosses) are both younger
and have a much larger extinction and/or infrared excess.
This last fact  is not surprising, since the IRAS data displayed 
in Figure 1 clearly show the concentration of dust around B30 \& B35.
However,  some of them might have  circumsubstellar
disks which could be responsible for part of the IR excess.
This assertion is supported by the Color-Color Diagrams
(CCD) displayed in Figure 7-8 (symbols as in Figure 3).
Many of the D\&M  members of B30 \& B35 are located close to
the Classical TTauri star locii. Moreover, most 
of them have strong H$\alpha$ emission, which might be due to
ongoing accretion from the circunstellar disk.

In these diagrams, we have overplotted the  isochrones
by Baraffe et al. (1998)  using 
the interstellar extinctions of
$A_V$=0.374, $A_R$=0.307, $A_I$=0.223, 
$A_J$=0.106, $A_H$=0.066, and $A_K$=0.04.
These values where computed from the
reddening measured by Diplas \& Savage (1994),
 E$(B-V)$=0.12, and  
the transformations by Rieke \& Lebofsky (1985).

\begin{figure}   
  \includegraphics[width=\columnwidth]{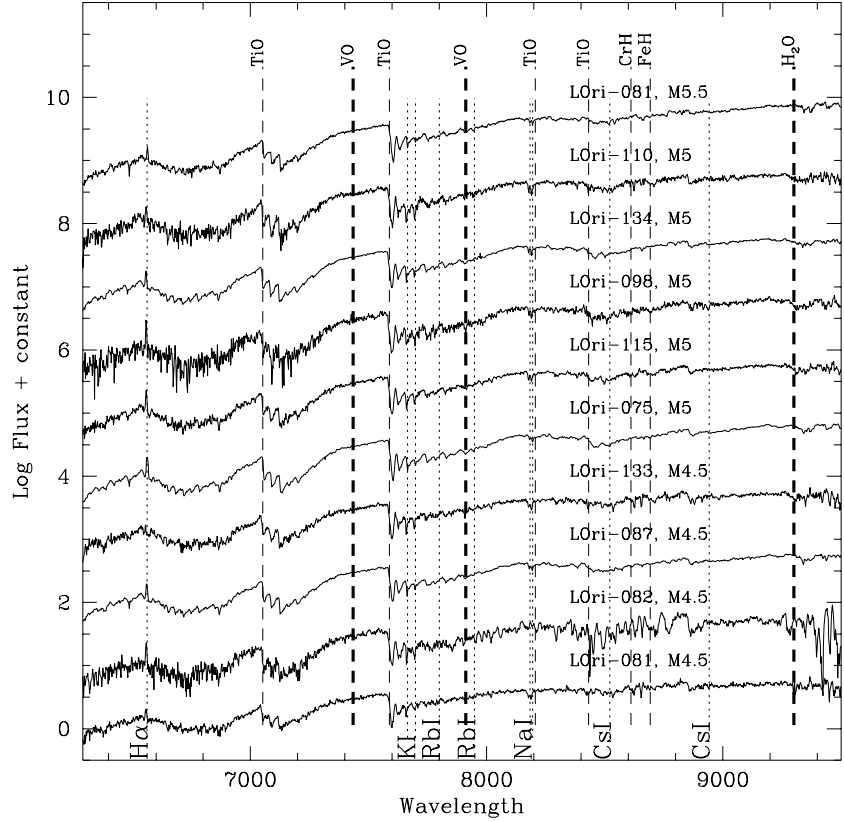}
  \caption{A sample of our low resolution spectra, taken with Keck/LRIS.
 Among other goals, these data were used for spectral classification
and membership confirmation. We have labeled  several relevant
 spectral features.
}
\end{figure}

\subsection{Follow-up: Low-resolution spectroscopy}

Our ongoing  comprehensive study of the  LOSFR 
also includes optical spectroscopy. We have obtained 
low-resolution spectra in November 3-5, 2002 at the Keck I telescope, 
using the LRIS spectrograph and the 400 l/mm grating,
achieving a resolution of R$\sim$1100 and a spectral
range of 6300-9600 \AA.
Later on, in March 9-11 2003, we obtained additional
spectral with the Magellan II telescope and the B\&C spectrograph.
In this case we used the 300 l/mm grating, the resolution was R$\sim$800,
and we covered the range  5000-10200 \AA.
Details can be found in Barrado y Navascu\'es et al. (2004a).
 In total, we observed 33 objects, both stellar and substellar,
with $I_C$ in the range 15.23 and  20.73 mag. Figure 9 includes some spectra.
We also observed several spectral templates and derived spectral types, 
which range from M4.5 to M7.5, which  allowed us to assess the 
membership status of this subsample more securely than using the 
purely photometric classification.

\begin{figure}   
  \includegraphics[width=\columnwidth]{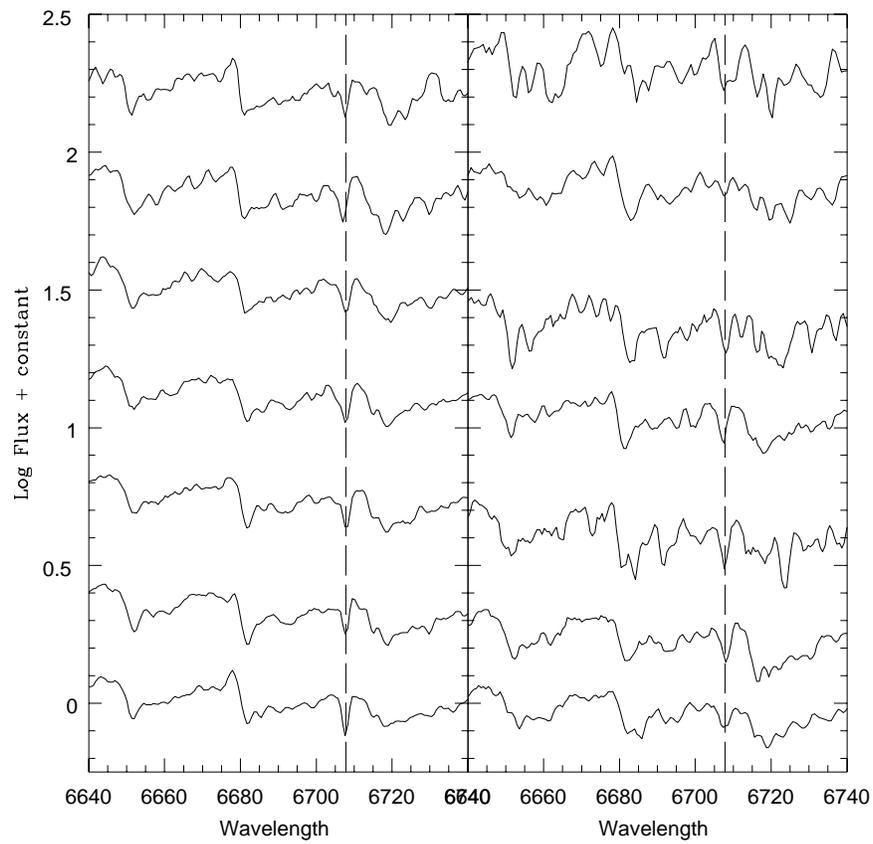}
  \caption{Detail around the lithium 6707.8 \AA{} for a sample of medium 
resolution spectrum.
}
\end{figure}

\subsection{Medium-resolution spectroscopy}

During the same campaigns, we also collected medium resolution spectra.
The details can be found in Barrado y Navascu\'es et al (2005, in preparation).
In both telescopes we used the 1200 l/mm grating, achieving a very similar 
resolution, R$\sim$3200, as measured with the Argon-Neon comparison lamps.
The spectra include the H$\alpha$ (6563 \AA) and lithium (6708 \AA) lines.
Figure 10 shows the spectral range around lithium for a subsample
of the objects observed at medium-resolution.
This alkali doublet is clearly seen in most of the objects, confirming beyond
any reasonable doubt the cluster membership.

\section{The properties}

\begin{figure}   
  \includegraphics[width=\columnwidth]{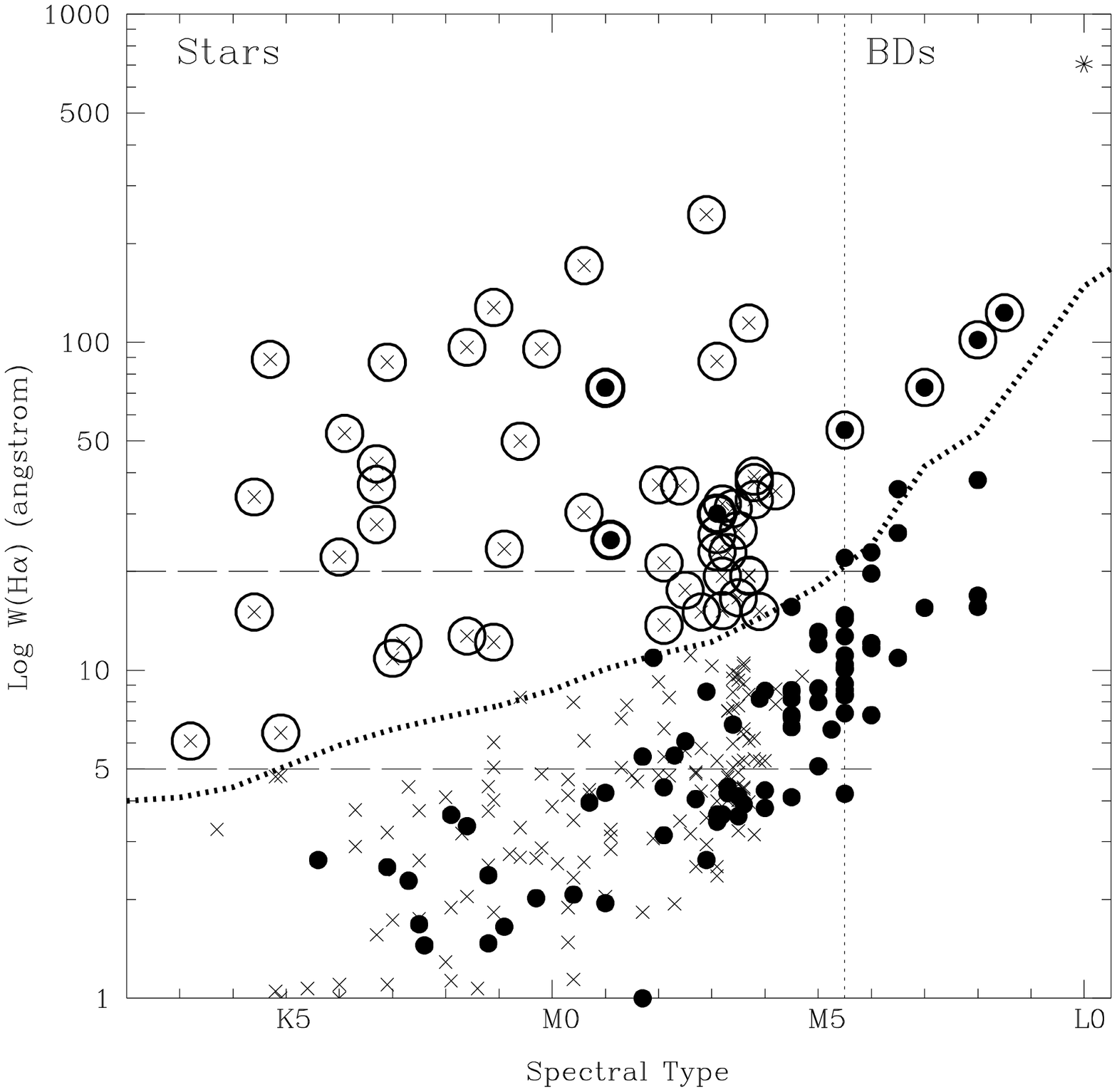}
  \caption{ 
H$\alpha$  equivalent width  versus  the spectral type.
We display data from D\&M as green crosses
for B30 \&B35 dark clouds and solid blue  circles for Collinder 69, 
 whereas our new data (CFHT/12K) for Collinder 69 
appear as  purple solid  circles.
The dotted line corresponds to the saturation criteria defined by
 Barrado y Navascu\'es \& Martin (2003), whereas the  two dashed 
lines are other criteria which have been used to define Classical
 TTauri stars. B30, B35 and Collinder 69 members with W(H$\alpha$) above the
saturation limit are highlighted with large open circles.
}
\end{figure}

\subsection{H$\alpha$ as a proxy of the accretion in Col~69, B30 \& B35}

We have measured the H$\alpha$ equivalent width  --W(H$\alpha$)--
 both in our low- and 
medium-resolution spectra of the Col~69 candidate members.
D\&M also have W(H$\alpha$) for the stars listed in their 
studies. Figure 11 displays these data versus the
derived spectral types. Symbols as in Figure 3.
The diagram also includes two horizontal segments 
(red, long-dashed lines), which corresponds to an ad hoc 
and widely used  criteria to discriminate between 
accretion and non-accreting stars
in SFRs. The red, dotted line is the accretion criterion 
proposed by Barrado y Navascu\'es \& Mart\'{\i}n (2003), which
is based on the saturation of the chromospheric
activity observed in young open clusters (30-125 Myr), which
appears at Log \{Lum(Halpha)/Lum(bol)\}=$-$3.3 dex.
Based on this criterion, we have classified the 
the three samples analyzed in this paper (our
own faint objects associated to Col~69 --purple, solid
circles--, the D\&M stars related to Col~69 --blue, solid
circles--, and the D\&M stars related to either B30 or B35
--green crosses--), into accretion  and non-accreting objects.
In the first case, we have over-plotted a large, open circle.
This classification has been shown in Figures 3-8.
 
The comparison between these three samples (actually two: the
objects related to Col~69, shown as purple and blue solid circles;
and those related to B30 or B35, shown as green crosses)
indicates that the fraction  of accreting objects is very different
in each group (a result already pointed out by D\&M).
In any case, the activity, coming from accretion or from the 
chromosphere, is much larger in B30 \& B35 than in Col~69.

\subsection{On the  age of Col~69}

\begin{figure}   
  \includegraphics[width=\columnwidth]{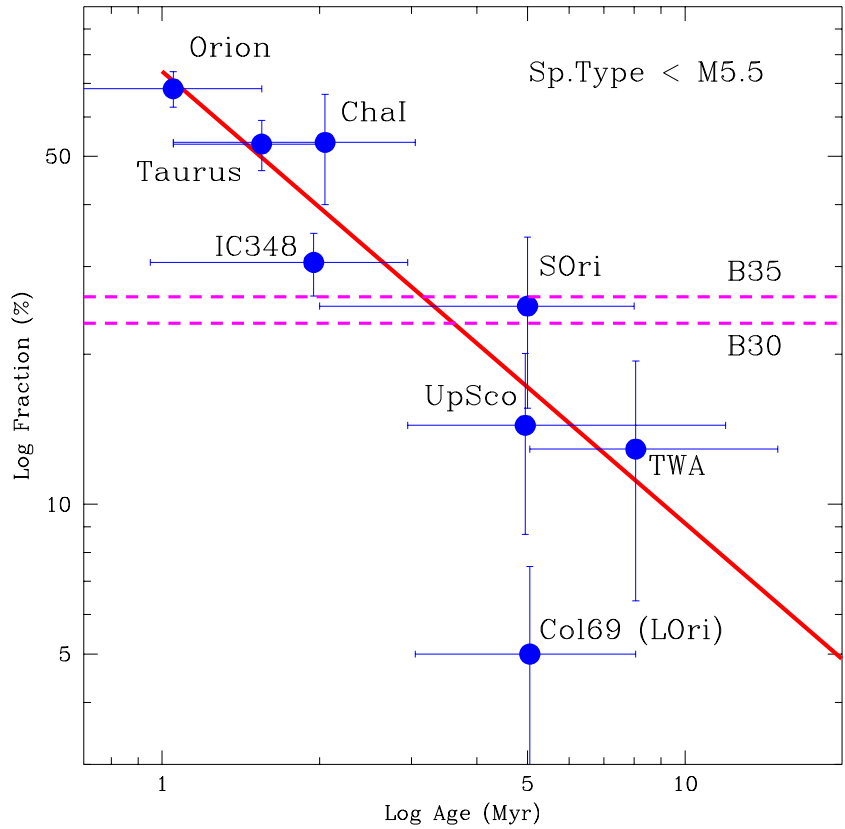}
  \caption{
Fraction of stars classified as  CTT stars  
(based on the H$\alpha$ equivalent width, see previous figure)
for different star forming regions and young clusters
(red solid circles).
The average trend between the fraction and the age is represented
with a red  thick, solid line.
The magenta,  dashed lines correspond to the  values for B30 \& B35,
}
\end{figure}

\subsubsection{An age estimate from the H$\alpha$}

We have compared the H$\alpha$ properties of Col~69 and B30/B35
with other SFRs and very young open clusters.
To do so, we have computed the ratio of stars
having  W(H$\alpha$) above the saturation criterion
with the total number of members with known H$\alpha$.
Figure 12 shows the results, and includes 
data from Orion, Chamaeleon I, Taurus, Sigma Orionis, 
Upper Scorpius and the TW Hydrae Association
(see details and references in 
Barrado y Navascu\'es \& Mart\'{\i}n 2003 and
Barrado y Navascu\'es, Mohanty and Jayawardhana 2005, in preparation).
As far as we know, there is no accurate age estimate
for B30 \& B35, but the clear trend observed in the other 
associations suggests that they are about 3 Myr, younger than
the age estimated by D\&M for Col~69.
However, the fraction of accreting stars of Col~69 (or
the fraction of very active members) is much lower than expected,
indicating that either Col~69 is much older than previously thought
(but the star $\lambda$$^1$ Orionis cannot be older than
about 7 Myr, since with a mass around 25 Myr would have 
exploded as a supernova), or some event has removed the 
disks around the Classical TTauri stars belonging to the 
association. The strong wind and UV flux from $\lambda$$^1$ Orionis
(O8 III spectral type)
might have played this role, at least in the case of the members
close to the central star. Or D\&M might have hit the answer when
they suggested that a supernova, more massive than $\lambda$$^1$ Orionis,
 went off in Col~69, inducing the  star formation  in B30 \& B35
 (i.e., they have to be younger 
that Col~69), and just initiated the process in LDN1588 and LDN1603
(Figure 1).

A third possibility would be that  $\lambda$$^1$ Orionis
is not related to Col~69, which is in unlikely based on the
measured distances and overall properties.

\begin{figure}   
  \includegraphics[width=\columnwidth]{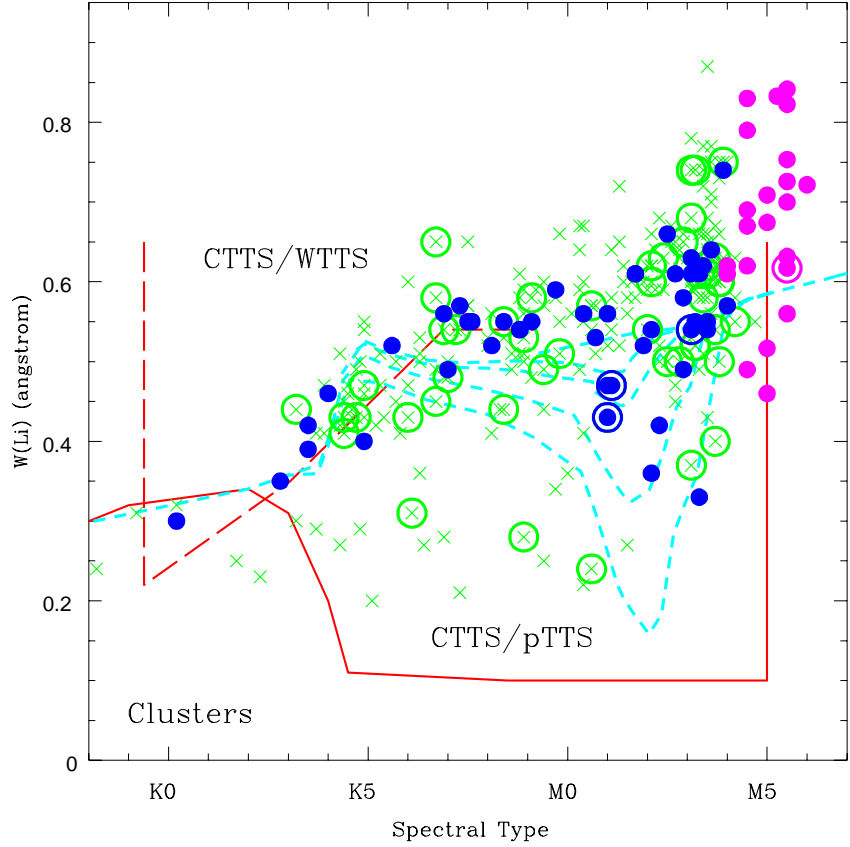}
  \caption{
Lithium equivalent width versus the spectral type.
We display data from D\&M as green crosses
for B30 \&B35 dark clouds and solid blue  circles for Collinder 69, 
 whereas our new data (CFHT/12K) for Collinder 69 
appear as  purple solid  circles.
The solid line corresponds to the upper envelope of the values
measured in young open clusters such as IC2391, IC2602, the Pleiades and M35.
The long-dashed line delimits the areas for weak-line and post-TTauri
stars (adapted from Mart\'{\i}n 1997 and Mart\'{\i}n \& Magazz\`u 1999).
The cyan, short-dashed lines correspond to the
Baraffe et al. (1998) lithium depletion isochrones for ages of 1 Myr
--cosmic abundance, A(Li)=3.1--, 8, 10, 15 and 20 Myr
(curves of growth from Zapatero Osorio et al$.$ 2002, 
effective temperature scale by Luhman 1999).
}
\end{figure}

\subsubsection{The age and lithium equivalent width}

Lithium, due to its fragility and easy destruction at low temperatures
inside the core of stars and high mass brown dwarfs,
is key to understand their internal structure and evolution.
Moreover, it can be used as an independent age indicator.
The theoretical models predict that no lithium depletion takes place at any 
mass during the first few million years ($\sim$3 Myr, see
D'Antona \& Mazzitelli 1994, 1997; Baraffe et al. 1998; Siess
et al. 2000). After that, a fast depletion takes places, especially 
for early M stars. This property has been used to determine,
 in a very accurate way, the age of IC2391, the Alpha Persei cluster,
and the Pleiades (Stauffer et al. 1998, 1999; Barrado y Navascu\'es
et al. 1999, 2004b). However, 
for younger ages, lithium  becomes a less accurate age indicator
because the lithium depletion chasm becomes shallower.  
More physically, lithium depletion becomes much harder to model
because the stars involved develop radiative cores, and
the amount of lithium observed in the photosphere depends 
critically on the exact point in the evolution of the star when the
radiative core develops and on any mechanism which may mix matter
across the core/envelope boundary (see Jeffries 2004).
Moreover, activity and spottedness may affect the ability to accurately
infer the effective temperature of the star and may affect the observed 
lithium equivalent width, making derivation of a lithium
abundance subject to error (Barrado y Navascu\'es
et al. 2001a; Stauffer et al. 2003).

In any case, we have compared the lithium equivalent widths
--W(Li)-- for our three samples.  Figure 13 displays W(Li)
versus the spectral type. Lithium depletion isochrones
--computed using models by Baraffe et al. (1998), curves of growth by
Zapatero Osorio et al. (2002) and the intermediate effective temperature
scale by Luhman (1999)-- have been included as cyan, short-dashed lines
(for ages of 1, 8, 10, 15 and 20 Myr).
The scatter of the W(Li) is very large, and we
cannot see any trend with accretion (large open circles), induced by the
veiling of the spectrum.
However, it seems that the upper limit for B30 \& B35 is
systematically  larger 
(about 0.1 \AA) than in the case of Col~69. Might this indicate
an enrichment of the material used to built up the members of these
two clouds? The alternative would be that some depletion --which seems
not to depend on the spectral type-- has already occurred in Col~69.
Moreover, 
the W(Li) distribution does not follow the lithium depletion isochrones
either, but they impose an upper limit to the age of about 20 Myr, which 
is not very helpful in any case.
Models by D'Antona \& Mazzitelli (1994, 1997)  or Siess et al. (2000)
do not fare any better.
 Additional theoretical work regarding the
Pre-Main Sequence lithium depletion has to be carried out.

\begin{figure}   
  \includegraphics[width=\columnwidth]{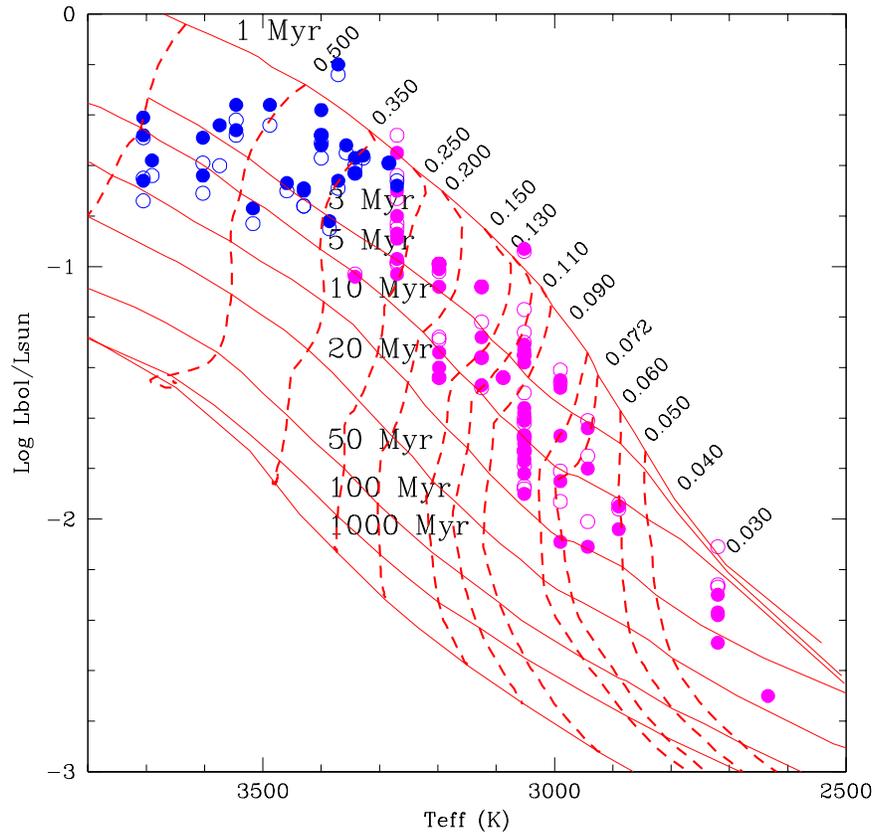}
  \caption{HR diagram for the Lambda Orionis cluster (Collinder 69).
Isochrones (red solid lines) and evolutionary tracks (red dashed lines) 
were computed by Baraffe et al. (1998).
 The age can be estimated as 3-10 Myr.
}
\end{figure}

\subsubsection{An age estimate from the HR diagram}

An age estimate for the Col~69 cluster can be obtained by locating 
its bona fide members in  a HR diagram (Figure 14).
Luminosities were derived  from either  the $Ic$ or the $Ks$
magnitudes --open and solid circles, respectively-- 
and bolometric corrections
by Comer\'on et al. (2000) and Tinney et al. (1993).
As usual, blue and purple colors denote stars and brown dwarfs
from D\&M and our own dataset from CFHT/12K.
The isochrones and evolutionary tracks come from Baraffe et al. (1998).
Effective temperatures have been derived using the
temperature scale by   Luhman  (1999) for
intermediate gravity. This figure indicates that 
the age of the cluster is bracketed by 3 and 10 Myr, allowing some 
room for binarity. 
Moreover, based on models by Schaller et al. (1992), 
 the star $\lambda$$^1$ Orionis, with a mass close to 25 $M_\odot$,
is about 6 Myr. The models impose a maximum age of  7 Myr, since
it would have become a supernova. Therefore, we adopt this value as the maximum
age for the Col~69 cluster.

\subsection{The Initial Luminosity and Mass Functions}

\begin{figure}   
  \includegraphics[width=\columnwidth]{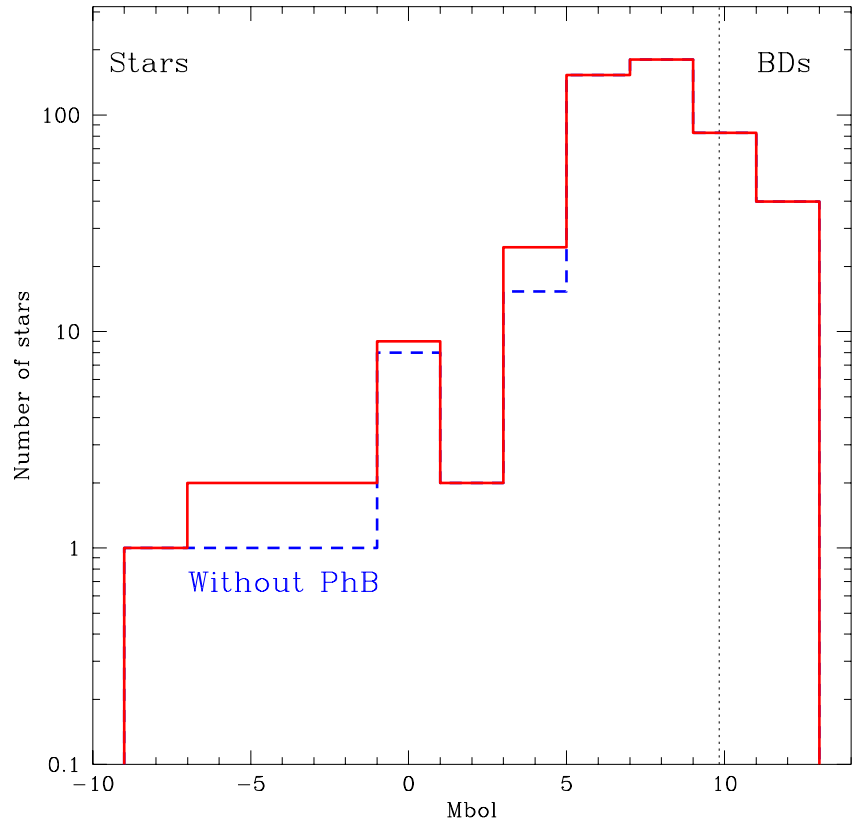}
  \caption{
Luminosity Function for the Collinder 69 cluster.
The blue dashed lines line was computed removing some possible members 
which  fall above the cluster locus, in   the photometric  sequence.
}
\end{figure}

\subsubsection{The Luminosity Function}

One of our goals is to derived the Initial Mass Function (IMF) in different
agglomerates of the LOSFR (B30 \& B35, Col~69, etc), and to establish
the effect of the different environmental conditions. So far, we have 
been able to derive the IMF for Col~69. The first step is to 
obtain the cluster Luminosity Function (LF). For the lower end, we
 have used a subsample of the Col~69 members from our survey 
with CFHT/12K which have been  classified as probable and possible members. 
For the middle range of masses, we have used the D\&M data which is located 
in the same projected area of the sky than our CFHT/12K data.
Finally, in order to extend the LF to the upper mass end, 
we have included the O, B, A and F stars listed in Murdin \& Penston (1997),
after removing some of them, which might not be cluster members because of their
location in CMDs. Additionally, we have included  an A7 star
taken from SIMBAD (HDE244927), which might be a member based on its photometric
characteristics (the Tycho parallax does not agree with membership,
 but errors are  quite large). 

The LF is illustrated in Figure 15. The bolometric magnitudes were derived 
in different ways: For the Murdin \& Penston (1977) data, we used the 
V magnitude, the cluster distance modulus and the bolometric corrections by 
Schmidt-Kaler (1982), which depend on the spectral type.
In the case of the D\&M and CFHT/12K datasets, we derived the
 bolometric magnitudes from the $I_C$ magnitude, the cluster distance modulus,
and the bolometric corrections by Comer\'on et al. (2000), which 
depend on the color $(R-I)$. The red solid line represents the LF when
all members (probable and photometric binaries) are considered. The blue
dashed line was computed after removing possible photometric binaries.
The black, dotted vertical segment 
distinguishes between the stellar and substellar domain.
The dip at about M(bol)=2 mag is an artifact due to the lack of completeness
for F stars. 

\begin{figure}   
  \includegraphics[width=\columnwidth]{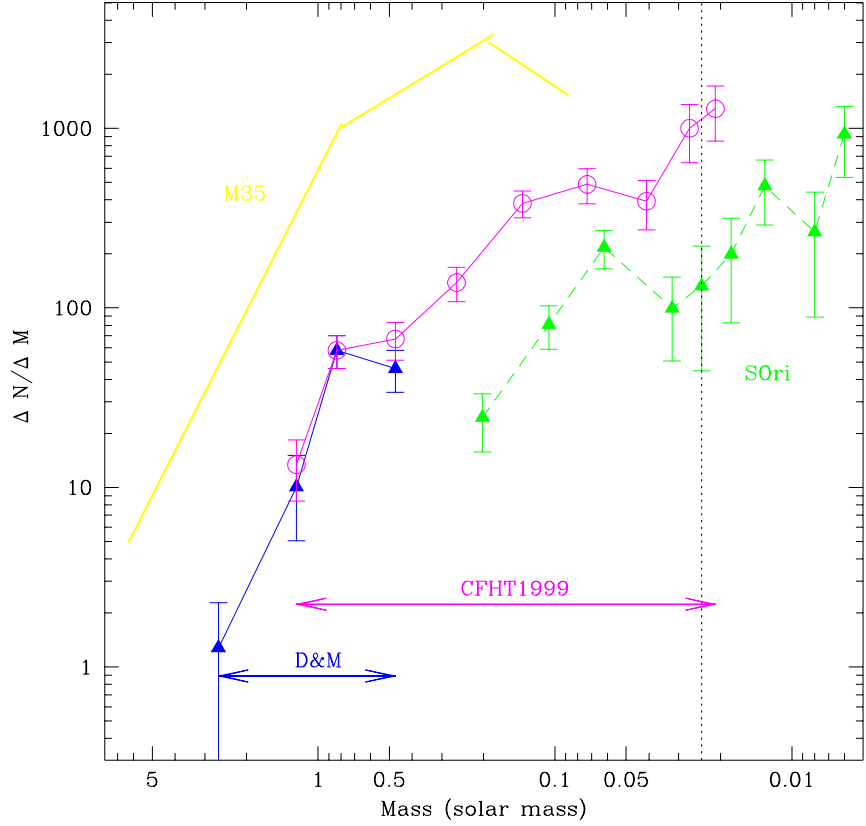}
  \caption{
Initial Mass Function for the Collinder 69 cluster 
(blue and purple lines), compared 
 with the older cluster M35 (150 Myr, yellow),
 where mass segregation has already
taken place, and the Sigma Orionis Cluster (5 Myr, green).
}
\end{figure}

\subsubsection{The substellar Initial Mass Function}

We have derived an IMF for the  low mass stars and brown dwarfs
based on the D\&M and CFHT/12K dataset. Although these two surveys
are not completely homogeneous, both provide $R_C$ and $I_C$ photometry
and allow the derivation of the LF and IMF in exactly the same way.
For this purpose, we have restricted ourselves to the 42$\times$28 
arcmin FOV of the CFHT/12K. The masses were computed from the $I_C$
magnitude, using theoretical models by Baraffe et al. (1998).
Additional details can be found in Barrado y Navascu\'es et al. (2004).
We used isochrones with different ages, but the results are
essentially independent of the considered age range. Therefore, we will
display only the results for 5 Myr, our assumed cluster age.  

Figure 16 includes the results for Col~69 (blue and purple lines, 
for D\&M and CFHT/12K, respectively), as well as a comparison
with the Sigma Orionis cluster (5 Myr, green dashed line) and 
M35 (150 Myr, yellow thick line).
The completeness limit for Col~69 is indicated by the 
black, dotted vertical segment.
In the case of M35, a very rich cluster (Barrado y Navascu\'es et al. 2001b
and references therein),
some mass segregation has taken place, as can be appreciated by the drop of 
the MF below 0.2 $M_\odot$. The Sigma Orionis IMF was computed for this paper
exactly in the same way as that of Col~69, based on the data published in the 
literature (see B\'ejar et al. 2001 and references therein).
The same is valid for M35.
Note the dip both in the case of Col~69 and Sigma Orionis cluster 
for masses around 0.04 $M_\odot$. This mass range corresponds, for 5 Myr, 
to $\sim$M7 spectral type. Dobbie et al. (2002) have found the same kind
of structure in older clusters such as IC2391, the Alpha Persei cluster,
 et cetera, at this spectral type (i.e., the same effective temperature
but  different masses, because of the evolution with age). They have 
interpreted  this fact as a consequence of 
a drop in the luminosity-mass relation, perhaps due to the
 the beginning of dust formation in the stellar/substellar atmospheres 
of objects at these temperatures.

We have fitted a power low to the derived IMF (rather, the 
Mass Spectrum, $dN/dM$ $\propto$ $M^{-\alpha}$).
The derived index is $\alpha$=+0.60$\pm$0.06 
across the stellar/substellar limit
(0.03-0.14 $M_\odot$), and a slightly steeper index $\alpha$=+0.86$\pm$0.05
 over the whole mass range from $\sim$0.024 $M_\odot$ to 0.86 $M_\odot$,
 using a 5 Myr isochrone.
An isochrone from Burrows et al. (1997) gives  $\alpha$=+0.69$\pm$0.17
in the range 0.20--0.015 $M_\odot$, whereas models from D'Antona \& Mazzitelli
(1997) are almost identical --regarding the power law index--
to those obtained with  Baraffe et al. (1998).  
 On the other hand, 3 and 10 Myr isochrones from Baraffe et al. (1998)  yield
 $\alpha$=+0.92$\pm$0.04 and  $\alpha$=+0.71$\pm$0.06, respectively
(again, in the range 0.024 $M_\odot$ to 0.86 $M_\odot$).
For mass in the range 3--1 $M_\odot$, the spectral index is very similar to
that of Salpeter.

 The slope of Lambda  Ori MF at lower
masses and into the substellar domain is quite similar to that derived for
other young clusters by some of us,
 e.g. Sigma Orionis ($\alpha$=+0.8, B\'ejar et
al. 2001), Alpha Per ($\alpha$=+0.6, Barrado y Navascu\'es et al. 2002)
and the Pleiades ($\alpha$=+0.6, Bouvier et al. 1998; Moraux et al. 2003). 
The age of these
clusters is estimated as 5, 80 and 125 Myr, respectively (Zapatero Osorio
et al. 2002; Stauffer et al. 1998, 1999; Barrado y Navascu\'es et
al. 2004). The $\alpha$ index is also similar to the results 
obtained in other stellar associations 
such as Trapezium, IC348 or Taurus (Luhman et al. 2000, 2003;
 Lucas \& Roche 2000; Hillenbrand \& Carpenter 2000;
 Najita et al. 2000; Preibisch et al. 2002; Brice\~no et al. 2002;
 Muench  et al. 2003).

\begin{figure}   
  \includegraphics[width=\columnwidth]{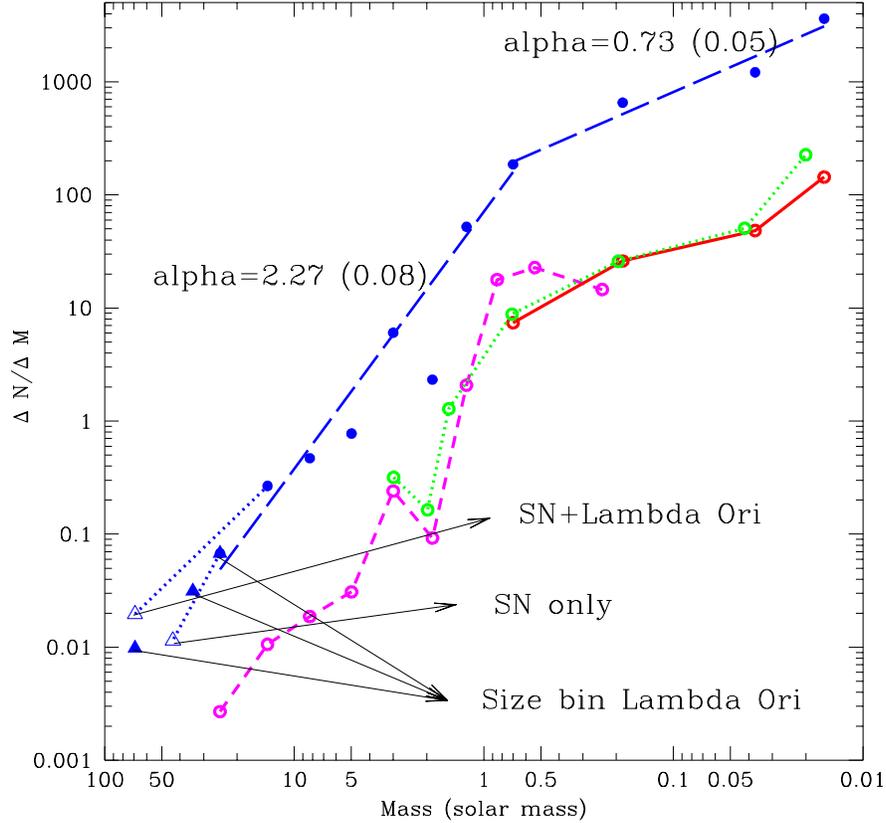}
  \caption{The Initial Mass Function of Collinder 69 cluster
in the range 50-0.02 $M_\odot$. The blue lines correspond to the
final IMF, whereas the purple, green and red lines were computed using 
several  models, valid for different mass ranges (see text).
}
\end{figure}

\subsubsection{A 50-0.02 $M_\odot$ Initial Mass Function}

Finally, we have derived an IMF for the whole cluster population 
down to 0.02 $M_\odot$. This has been achieved by using the LF presented 
in section  3.3.1 (Figure 15). However, since no theoretical
set of evolutionary tracks considers the whole mass range of our
data, we have been forced to use three different models:
For high mass stars we have considered the isochrones
by Girardi et al. (2002), whereas for the low mass end those  
of Baraffe et al (1998). To guarantee that the results  merge 
nicely and there is no obvious bias in the results, we also used the models
by D'Antona \& Mazzitelli (1997), which cover the masses in between 
the first two.  The results are displayed in Figure 17, where
we have plotted the IMFs derived with the models by Girardi et al. (2002), 
D'Antona \& Mazzitelli (1997), and Baraffe et al (1998) as purple dashed,
 green dotted and red solid lines, respectively. The final result
is illustrated by the blue, long-dashed line.
Note that this IMF has two important differences in its methodology with
the one computed is section 3.3.2. First, in this case we have used the 
bolometric magnitudes  instead the $I_C$ magnitudes. Second, the binning 
of the magnitudes  is  wider in this case, in order to diminish the
statistical errors (specially in for high masses, were the number 
of objects is small). As noted before, the dip at 2 $M_\odot$ is
due to the lack of completeness for F stars.

\begin{figure}   
  \includegraphics[width=\columnwidth]{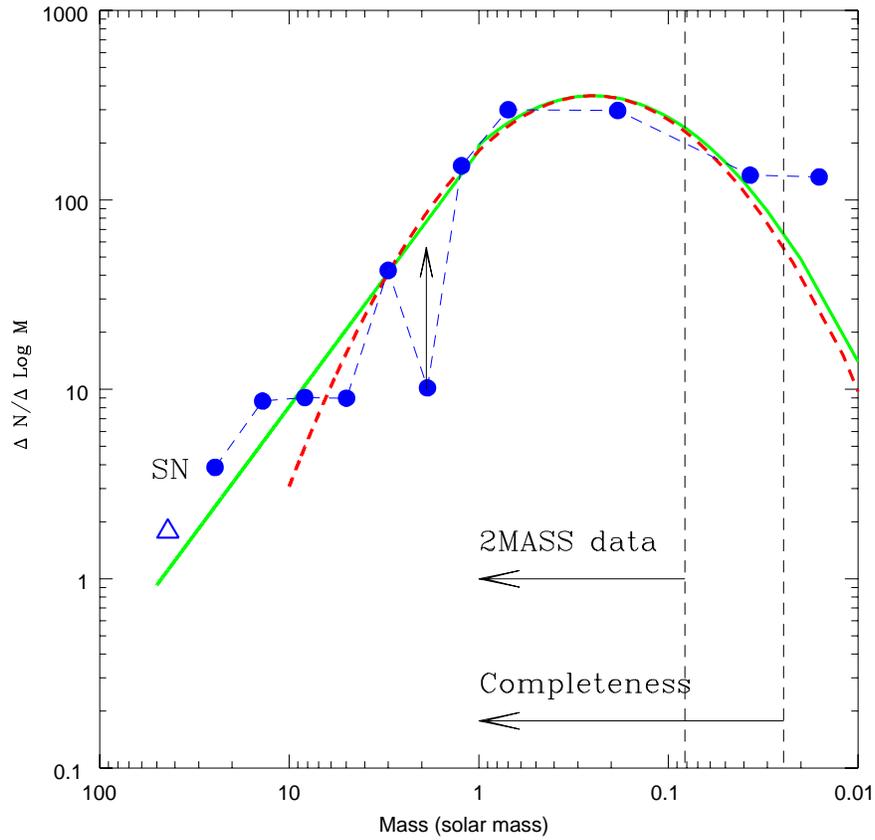}
  \caption{The Initial Mass Function, in log-form, of Collinder 69 cluster
in the range 50-0.02 $M_\odot$ (blue solid circles). The blue empty 
triangle represents the possible supernova.
The green-solid and red-dashed lines show  the fits by Chabrier (2003, 2004)
for the young disk and by Moraux et al. (2003) for the Pleiades.
}
\end{figure}

In order to compute the IMF at the high mass end, we have assumed
 different possibilities. On one hand, we have considered
that the most massive star  in the cluster is $\lambda$$^1$ Orionis, 
but we have assumed that the corresponding mass 
bin had different values in its 
upper side (33, 50 and 120 $M_\odot$), with different widths and mass averages.
These three possibilities are located in the figure as blue, solid triangles.
On the other hand, we have considered the possibility that Col~69
harbored in the past a more massive star, which  exploded about 1-2 Myr ago
as a supernova, as postulated by D\&M. In this case, we can compute the IMF
assuming that the SN and the star  $\lambda$$^1$ Orionis fall in the same and
in different  mass bins. They are represented as blue, open triangles.
Regardless the high mass end of the IMF, the slope of the IMF 
has an index of $\alpha$=+2.27$\pm$0.08 (20--0.8 $M_\odot$),
whereas the value is much smaller for lower masses, 
$\alpha$=+0.73$\pm$0.05 (0.8-0.02  $M_\odot$).
The overall trend is very similar to the more massive and 
older M35 cluster (Barrado y Navascu\'es et al. 2001b).

We have also compared the Col~69 IMF with the Pleiades 
and the young disk population, in log-form. 
They are depicted in Figure 18, as blue, dashed-red 
and solid green, respectively. In order to computed the last two 
Mass Functions, we have used the expression by Chabrier (2003):

\begin{eqnarray}
\xi(\log \,M)&\propto& \exp\Bigl\{-{(\log \, M\,\,-\,\,\log \, Mc)^2\over 2\times \sigma^2}\Bigr\},\,\,\, m\le 1\,M_\odot \nonumber \\
\xi(\log \,M)&\propto& M^{-1.35}\,\,\,\,\,\,\,\,\,\,\,\,\,\,\,\,\,\,\,\,\,\,\,\,\,\,\,\,\,\,\,\,\,\,\,\,\,\,\,\,\,\,\,\,\,\,\,\,\,\,\,\,\,\,\,\,\,\, , m\ge    1\,M_\odot \nonumber  
\label{IMF1}
\end{eqnarray}

\noindent where $Mc$=0.25 and 0.20,  and $\sigma$=0.52 and 0.55,
 for the Pleiades and the young disk population
(after Moraux et al. 2003 and Chabrier 2004, 
respectively). As can be seen, these IMFs are very similar,
 except for the fact that Col~69 is not complete for some
mass ranges, and the Pleiades, due to its age (about 125 Myr),
 does not contain  massive stars.

\section{Conclusions}

We are conducting an ambitious study of the star formation 
and the properties of the LOSFR, a wide area which includes 
several dark clouds and clusters. In particular, we have collected
deep optical photometry, low-  and medium resolution spectroscopy,
and mined the 2MASS database to select a sample of bona fide members of
 the Col~69 cluster, which is associated to the star $\lambda$$^1$ 
Orionis, located in the center of the LOSFR.
With this sample and previously, brighter data, we have been able to derive 
the Luminosity and Initial Mass Functions of the Col~69 cluster,
in the mass range 50-0.02 $M_\odot$. Two different behaviors are
evident, with a turning point about 0.8  $M_\odot$. The high mass 
stars follow a pattern similar to a Salpeter's index, whereas
low mass stars and brown dwarfs  have a much shallower index.
In any case, the IMF keeps growing well within the substellar domain
(although this is not the case for the LF).
Additionally, the IMF, in a log-form, is very 
similar to the Pleiades and the 
young disk population.

 On the other hand,
we have also studied the properties of the Col~69 members (H$\alpha$
emission, lithium equivalent width, CMDs and CCDs, HR diagram).
We have been able to derive an age for the cluster 5$\pm$2 Myr.
Moreover, when comparing these data with members of the B30 \& B35 dark 
clouds, we have inferred that these two associations
have a much larger population of Classical TTauri stars
(both from the CCD and the distribution of the H$\alpha$ emission of 
their members), are much younger (with an age of about 3 Myr)
and have a similar distribution of the lithium equivalent width
for the K5-M5 members.


\end{document}